\documentclass[aps,prl,twocolumn,titlepage,nofootinbib]{revtex4}
\usepackage{graphicx,physics}% Include figure files
\usepackage{color}
\usepackage{dcolumn}% Align table columns on decimal point
\usepackage{latexsym}

\usepackage[normalem]{ulem}
\usepackage{hyperref,amssymb}
\usepackage{url}
\usepackage{color,soul}
\usepackage{graphicx}
\usepackage{verbatim}
\usepackage{multirow}
\usepackage{amsmath}
\newcommand{\beq}{\begin{eqnarray}}
\newcommand{\eeq}{\end{eqnarray}}
\usepackage{mathrsfs}
\usepackage{float,soul}
\usepackage[dvipsnames]{xcolor}
\usepackage{mathtools}
\usepackage{slashed}
\usepackage{physics}	% installed packages for typing maths and physics
\usepackage{graphicx}   % need for Fig.ures
\usepackage{epstopdf}
\usepackage{subfigure}  % use for side-by-side Fig.ures
\usepackage{hyperref}   % use for hypertext links, including those to external documents and URLs
\usepackage{bbold}
\usepackage{wasysym}
\usepackage{feynmp}
\usepackage{hyperref}
\hypersetup{colorlinks,
}
\usepackage{bm}%varepsilon

\usepackage[latin1]{inputenc}

\begin{document}

\title{Impact of elastic inhomogeneity on collective dynamical properties investigated by field theoretical description in real space}
\author{Cunyuan Jiang$^{1,2}$}
\email{cunyuanjiang@sjtu.edu.cn}
\address{$^1$School of Physics and Astronomy, Shanghai Jiao Tong University, Shanghai 200240, China}
\address{$^2$Wilczek Quantum Center, School of Physics and Astronomy, Shanghai Jiao Tong University, Shanghai 200240, China}

\begin{abstract}
     Interpreting the vibrational properties of amorphous solids beyond Debye's theory is challenging due to the presence of inhomogeneity on the mesoscopic scale. In this work, we model this inhomogeneity by real-space fluctuating elasticity with a spatially correlated distribution and calculate the dynamical properties using an exact real-space field theoretical approach. Our results clarify that the excess low-frequency density of states (DOS) originates from a selective scattering effect (stronger scattering of short wavelengths) induced by elastic inhomogeneity. The visualization of the local DOS in real space reveals the existence of anomalous modes, highly excited spots, at low frequencies. The findings regarding these highly excited spots and the selectivity of the correlation length were missed in previous perturbative field approaches in wave-vector space, and they align with recent progress from particle-level simulations and experiments. These results provide concrete insights into the low-frequency vibrational anomaly of amorphous solids from the perspective of simple elastic inhomogeneity.
\end{abstract}
\maketitle

\section{Introduction}
The collective dynamical properties of crystals can be interpreted within Debye's theory, which is based on translational symmetry and its spontaneous breaking.\cite{debye} This paradigm is challenged when applied to amorphous solids due to the presence of mesoscopic scale inhomogeneity inherent to amorphous materials. The observation of anomalous vibrational density of states (DOS) and specific heat in the low energy region, which violate the predictions of Debye's theory and are commonly referred to as the Boson peak (BP) in the literature, in various amorphous solids suggests the existence of a mechanism describing the competition between homogeneity and inhomogeneity, giving rise to collective dynamical anomalies.\cite{PhysRevB.4.2029,PhysRevLett.112.165901,PhysRevB.67.094203,PhysRevLett.133.188302,PhysRevResearch.6.023053,Edan2023,Gj2023}

Previously, the inhomogeneity of amorphous solids has been modeled by inhomogeneous elasticity, which assumes a spatially fluctuating elastic constant, using a field theoretical description.\cite{PhysRevLett.100.137402,PhysRevLett.98.025501} Within the perturbative approach, where the impact of elastic inhomogeneity is treated as a perturbation to the field theoretical propagator of plane waves, the low frequency anomalous DOS can be reproduced phenomenologically. However, the perturbative approach has been questioned for underestimating the role of microscopic interactions.\cite{PhysRevLett.123.055501,PhysRevLett.130.236101,Shintani2008,Tong2015} On the other hand, the field theoretical description in wave vector space cannot provide information about real space processes, which have been rapidly developing with the help of simulations and experiments.

Through dynamical simulations\cite{Hu2022,PhysRevLett.127.215504,10.1063/1.4769267,Zhang2011,Wang2019,Corrado2020,10.21468/SciPostPhys.15.2.069,Xu2024} and particle level experiments on granular matter\cite{PhysRevLett.133.188302,PhysRevB.98.174207,Zhang2017} and colloidal systems,\cite{sun2025disentanglinghierarchicalrelaxationsglass,PhysRevLett.108.095501} the collective dynamical anomaly in amorphous systems has been associated with localized modes. These modes are spatially limited (unlike extended plane waves) and temporally long lasting, corresponding to low frequencies. Localized modes are therefore considered to be the origin of the anomalous low frequency DOS in amorphous solids. In addition to identifying the existence of localized modes, the results of particle level investigations have also confirmed that the frequency of the anomalous DOS is inversely proportional to the size of the localized modes, i.e., \(\omega_{BP} \sim v/l\), where \(v\) is the speed of sound and \(l\) is the characteristic size. This size dependence of where the anomalous DOS appears indicates the non phononic nature of the vibrational anomaly. Theories of phonon coupling with localized modes have been developed to model this size dependence within the wave vector space perturbative approach.\cite{Jiang_2024,PhysRevB.101.174311} However, the origin of localized modes, vibrational inhomogeneity, and their visualization in real space for comparison with simulations and experiments remain missing and out of reach. For inhomogeneous elasticity in real space, although the perturbative theoretical framework in wave vector space has inevitable limitations, it remains promising due to its simplicity in real space for modeling the mesoscopic scale inhomogeneity of amorphous solids in the continuum limit, and thus for investigating the collective dynamical anomalies induced by inhomogeneity.

In this work, we present the results of a real space field description of the collective dynamical properties of an inhomogeneous elastic medium. First, the difficulties of the field description in wave vector space when encountering elastic inhomogeneity are discussed analytically, to be compared with the numerical simplicity of the real space field description. Then, the emergence of anomalous low frequency DOS is demonstrated when elastic inhomogeneity is introduced. Furthermore, it is shown that the frequency at which the anomalous DOS appears changes inversely with the correlation length of the elastic inhomogeneity, as confirmed by simulations and experiments. Finally, it is demonstrated that emergent vibrational inhomogeneity is a direct result of elastic inhomogeneity, characterized by associated low frequency highly excited spots at soft regions. The results confirm that two features, correlation length dependence and the emergence of highly excited spots, can be captured by the model of inhomogeneous elasticity as long as the field description is constructed in real space instead of using the perturbative approach in wave vector space. These results provide insights into the origin of low frequency anomalous DOS and real space vibrational inhomogeneity in amorphous solids.

\section{Scalar inhomogeneous wave equation and its Green's functions}
The scalar wave equation in an inhomogeneous elastic medium has proven useful for investigating the dynamical properties of inhomogeneous systems.\cite{Jiang_2025,PhysRevLett.98.025501,PhysRevLett.100.137402} In 3D space, the equation reads:
\begin{equation}
    \partial_t^2 u(\boldsymbol{r}, t) - \nabla \cdot \left[\kappa (\boldsymbol{r}) \nabla u(\boldsymbol{r}, t) \right] = 0, \label{equationall}
\end{equation}
where \(\kappa (\boldsymbol{r})\) is the spatial distribution of the elastic modulus, and \(u(\boldsymbol{r}, t)\) represents the vibrational intensity at position \(\boldsymbol{r}\) and time \(t\). Applying the Fourier transform yields the equation in the frequency domain:
\begin{equation}
\begin{split}
    \omega^2 U_{\omega} (\boldsymbol{r}) + \nabla \cdot \left[\kappa (\boldsymbol{r}) \nabla U_\omega (\boldsymbol{r}) \right] &= 0, \\
    \left[\omega^2 + \kappa (\boldsymbol{r}) \nabla^2 + \nabla \kappa (\boldsymbol{r}) \cdot \nabla\right] U_{\omega} (\boldsymbol{r}) &= 0 ,
\end{split}
\label{frequencyequation}
\end{equation}
where
\begin{equation}
    U_\omega (\boldsymbol{r}) = \int dt \, u(\boldsymbol{r}, t) e^{-i \omega t}
\end{equation}
is the Fourier transform of \(u(\boldsymbol{r}, t)\). To solve Eq.~\eqref{frequencyequation} for a given frequency, only boundary conditions are required. The Green's function of Eq.~\eqref{frequencyequation}, denoted \(G_{\omega} (\boldsymbol{r}, \boldsymbol{r}')\), satisfies:
\begin{equation}
\begin{split}
    \left[\omega^2 + \kappa (\boldsymbol{r}) \nabla^2 + \nabla \kappa (\boldsymbol{r}) \cdot \nabla\right] G_{\omega} (\boldsymbol{r}, \boldsymbol{r}') = - \delta (\boldsymbol{r}-\boldsymbol{r}'). \label{freenequation}
\end{split}
\end{equation}
Here, the differential operator \(\nabla\) acts only on \(\boldsymbol{r}\).

Because \(\kappa (\boldsymbol{r})\) varies in real space, applying the Fourier transform to Eq.~\eqref{frequencyequation} to convert it to the wave vector domain introduces convolutions that complicate the analysis. After Fourier transforming Eq.~\eqref{frequencyequation}, one obtains:\cite{Jiang_2025}
\begin{equation}
\begin{split}
    &\omega^2 \Tilde{G}_{\omega,\boldsymbol{k}}  - \Tilde{\kappa} (\boldsymbol{k}) \star k^2 \Tilde{G}_{\omega,\boldsymbol{k}} \\ 
    &- \boldsymbol{k} \Tilde{\kappa} (\boldsymbol{k}) \star \boldsymbol{k} \Tilde{G}_{\omega,\boldsymbol{k}} = -1 ,\\
    &\omega^2 \Tilde{G}_{\omega,\boldsymbol{k}}  - \int d \boldsymbol{q} \, \Tilde{\kappa} (\boldsymbol{k} - \boldsymbol{q}) q^2 \Tilde{G}_{\omega,\boldsymbol{q}}  \\ 
    &- \int d \boldsymbol{q} \, (\boldsymbol{k} - \boldsymbol{q}) \Tilde{\kappa} (\boldsymbol{k} - \boldsymbol{q}) \cdot \boldsymbol{q} \Tilde{G}_{\omega,\boldsymbol{q}} = -1 .
\end{split}\label{kgreen}
\end{equation}
Here, \(\star\) denotes convolution, and
\begin{equation}
    \begin{split}
        \Tilde{\kappa}_{\boldsymbol{k}} &= \dfrac{1}{2 \pi} \int d \boldsymbol{r} \, e^{-i\boldsymbol{k}\cdot \boldsymbol{r}} \kappa (\boldsymbol{r}) , \\
        \Tilde{G}_{\omega,\boldsymbol{k}} &= \dfrac{1}{2 \pi} \int d \boldsymbol{r} \, e^{-i\boldsymbol{k}\cdot \boldsymbol{r}} G_\omega(\boldsymbol{r},0) ,
    \end{split}
\end{equation}
are the Fourier transforms. Due to the convolutions, the Green's function equation in wave vector space becomes an integral equation that generally lacks an analytical solution. For a homogeneous elastic medium, \(\kappa (\boldsymbol{r}) = \kappa\) is constant, and its Fourier transform \(\Tilde{\kappa}_{\boldsymbol{k}} \sim \delta(0)\) is a delta function. Consequently, Eq.~\eqref{kgreen} simplifies to:
\begin{equation}
    \omega^2 \Tilde{G}_{0,\omega,\boldsymbol{k}}  - \kappa k^2 \Tilde{G}_{0,\omega,\boldsymbol{k}}  = -1 .
\end{equation}
The Green's function in wave vector space can then be solved analytically as:
\begin{equation}
    \Tilde{G}_{0,\omega,\boldsymbol{k}} = \dfrac{1}{-(\omega + i 0^+)^2 + \kappa k^2}. \label{green0}
\end{equation}
Here, \(0^+\) is a small positive number ensuring causality during the analytical continuation of the frequency into the complex plane. The Green's function in Eq.~\eqref{green0} describes free phonons in a homogeneous medium,\cite{PhysRevLett.100.137402,PhysRevLett.98.025501,PhysRevLett.122.145501} and its pole yields the linear dispersion relation \(\omega = \kappa^{1/2} k\).

As shown by Eq.~\eqref{kgreen} and Eq.~\eqref{green0}, the presence of inhomogeneity significantly complicates the Green's function. In particular, the Green's function in wave vector space is no longer local (determined by a single wave vector) but becomes extended (dependent on all wave vectors) due to the convolution. In contrast, addressing Eq.~\eqref{frequencyequation} directly in real space, while still not analytically solvable in general, allows for numerically exact solutions without resorting to perturbative approximations as required in wave vector space.\cite{PhysRevLett.98.025501,PhysRevLett.100.137402} Real space approaches also facilitate visualization and direct comparison with recent particle scale investigations, such as those examining the local density of states and flat dispersion relations.\cite{PhysRevLett.133.188302,Hu2022}

In real space, the Green's function \(G_{\omega} (\boldsymbol{r}, \boldsymbol{r}')\) for Eq.~\eqref{frequencyequation} can be obtained by numerically solving:
\begin{equation}
\begin{split}
    \left[(\omega + i 0^+)^2 + \kappa (\boldsymbol{r}) \nabla^2 + \nabla \kappa (\boldsymbol{r}) \cdot \nabla\right] \\ \times G_{\omega} (\boldsymbol{r}, \boldsymbol{r}') = - \delta (\boldsymbol{r} - \boldsymbol{r}'). \label{greenreal}
\end{split}
\end{equation}
The spectral function is then given by the imaginary part:
\begin{equation}
    A_\omega (\boldsymbol{r}, \boldsymbol{r}') = - \dfrac{1}{\pi} \text{Im} \, G_\omega (\boldsymbol{r}, \boldsymbol{r}'). \label{realspectra}
\end{equation}
The local density of states (DOS) corresponds to the spectral function when the excitation and observation points coincide, \(\boldsymbol{r} \equiv \boldsymbol{r}'\), and is thus \(A_\omega (\boldsymbol{r}, \boldsymbol{r})\). The total DOS is obtained by integrating the local DOS over real space:
\begin{equation}
    g(\omega) = 2 \omega \int d \boldsymbol{r} \, A_{\omega} (\boldsymbol{r}, \boldsymbol{r}). \label{dosreal}
\end{equation}
To study collective properties such as the dispersion relation, the spectral function in Eq.~\eqref{realspectra} can be expressed in wave vector space via Fourier transform:
\begin{equation}
    A(\omega, \boldsymbol{k}) = \int d\boldsymbol{R} \, A_{\omega}(\boldsymbol{R}) e^{i \boldsymbol{k} \cdot \boldsymbol{R}},
\end{equation}
where \(\boldsymbol{R} = \boldsymbol{r} - \boldsymbol{r}'\). Integrating over all directions:
\begin{equation}
    A(\omega, q) = \int d\boldsymbol{k} \, A(\omega, \boldsymbol{k}) \delta(q - |\boldsymbol{k}|),
\end{equation}
the DOS can also be written as:
\begin{equation}
    g(\omega) = 2 \omega \int dq \, A(\omega, q). \label{dosq}
\end{equation}
Using the finite difference method, Eq.~\eqref{greenreal} can be discretized into a system of linear equations \(H G_{\omega, \boldsymbol{r}, \boldsymbol{r}' = 0} = \delta_{\boldsymbol{r},\boldsymbol{r}' = 0}\), which can be solved efficiently. Here,
\begin{equation}
\begin{split}
    H &= (\omega + i 0^+)^2 I + \kappa_{\boldsymbol{r}} (D_x^2 + D_y^2) \\ 
    &+ ((D_x \kappa_{\boldsymbol{r}}) D_x + (D_y \kappa_{\boldsymbol{r}}) D_y ),
\end{split}
\end{equation}
represents the differential operator of Eq.~\eqref{greenreal} in finite difference form, where \(I\) is the identity matrix, \(\kappa_{\boldsymbol{r}}\) is the vector representation of \(\kappa (\boldsymbol{r})\), and \(D_{x(y)}\) are matrix representations of \(\partial_{x(y)}\).

\section{Impact on collective dynamical properties}
To investigate the impact of elastic inhomogeneity, the inhomogeneous elastic modulus \(\kappa (\boldsymbol{r})\) must be introduced. This distribution is generally characterized by two features. The first is the intensity of inhomogeneity, quantified by the maximum deviation of \(\kappa (\boldsymbol{r})\) from its average value \(\bar{\kappa}\), i.e., \(\text{Max}[|\kappa (\boldsymbol{r}) - \bar{\kappa}|]\). The second is the correlation length of the fluctuation in \(\kappa (\boldsymbol{r})\), which defines a length scale within which \(\kappa (\boldsymbol{r})\) does not vary significantly. For numerically solving Eq.~\eqref{greenreal} and computing the spectral function and VDOS, a feasible method to generate a length-correlated distribution of \(\kappa (\boldsymbol{r})\) is to apply a filter to \(\Tilde{\kappa}_{\boldsymbol{k}}\) in wave space:
\begin{equation}
\begin{split}
    \kappa(\boldsymbol{r}) = \int d \boldsymbol{k} \, \delta \left(|\boldsymbol{k}| - \dfrac{1}{L} \right) \\ \times \left[ \int d \boldsymbol{r} \, R(\boldsymbol{r}) e^{-i \boldsymbol{k} \cdot \boldsymbol{r}} \right] e^{i \boldsymbol{k} \cdot \boldsymbol{r}} .
\end{split}
\end{equation}
Here, \(R(\boldsymbol{r})\) is a random number generator that produces uncorrelated random numbers in real space, and \(L\) is the correlation length.

\begin{figure*}[htbp]
    \centering
    \includegraphics[width=\linewidth]{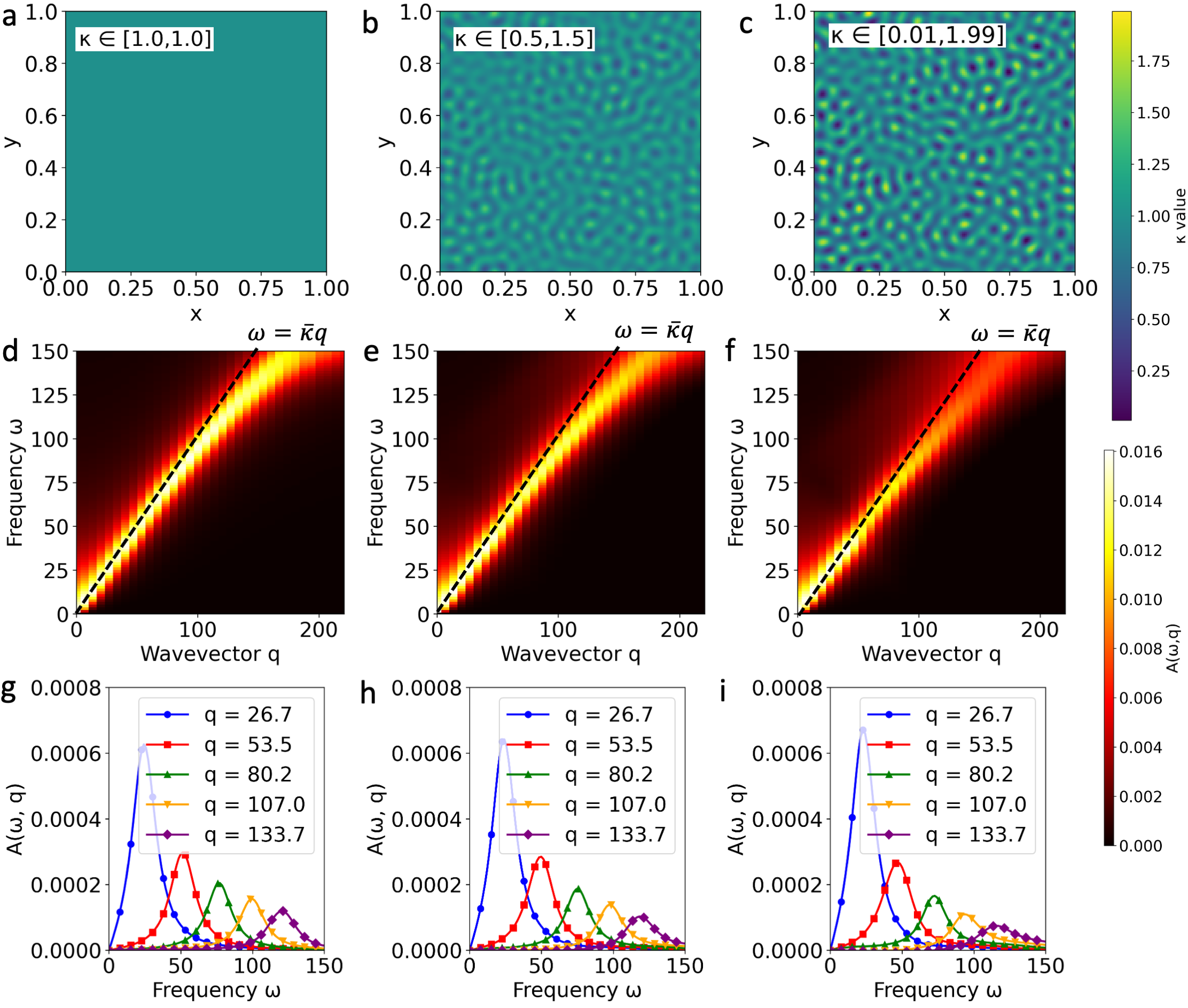}
    \caption{Real space distribution of \(\kappa (\boldsymbol{r})\) with fluctuation intensities of \(0\) (homogeneous) \textbf{a}, \(0.5\) (inhomogeneous) \textbf{b}, and \(0.99\) (inhomogeneous) \textbf{c}. The mean value is \(\bar{\kappa} = 1\), and the correlation length is \(0.01\). Spectral functions \(A(\omega, q)\) are shown in \textbf{d}, \textbf{e}, and \textbf{f}, corresponding to the inhomogeneity levels in \textbf{a}, \textbf{b}, and \textbf{c}, respectively. Cuts of the spectral function \(A(\omega, q)\) at various wave vectors \(q\) are shown in \textbf{g}, \textbf{h}, and \textbf{i}, for the same cases.}
    \label{figintensity}
\end{figure*}

Fig.~\ref{figintensity} a--c show the distribution of \(\kappa (\boldsymbol{r})\) in 2D space with the same correlation length \(L = 0.01 N\) (where \(N\) is the system size) but different inhomogeneity intensities. The corresponding dispersion relations and spectral functions are shown in Fig.~\ref{figintensity} d--f and g--i, respectively. The linear dispersion relation holds at small wave vectors for all inhomogeneity levels. At large wave vectors, the dispersion relation bends downward, approaching the Van Hove singularity induced by numerical discretization. Comparing the spectral functions \(A(\omega, q)\) for homogeneous and inhomogeneous cases (Fig.~\ref{figintensity} d--f and g--i), the impact of inhomogeneity can be categorized into two regimes. For wave vectors \(q\) smaller than \(1/(2L)\), the spectral function height is enhanced by inhomogeneity; a larger inhomogeneity intensity \(\text{Max}[|\kappa (\boldsymbol{r}) - \bar{\kappa}|]\) results in a higher spectral function. Conversely, for \(q > 1/(2L)\), the spectral function is suppressed, with greater inhomogeneity leading to lower spectral heights.

\begin{figure}[htbp]
    \centering
    \includegraphics[width=\linewidth]{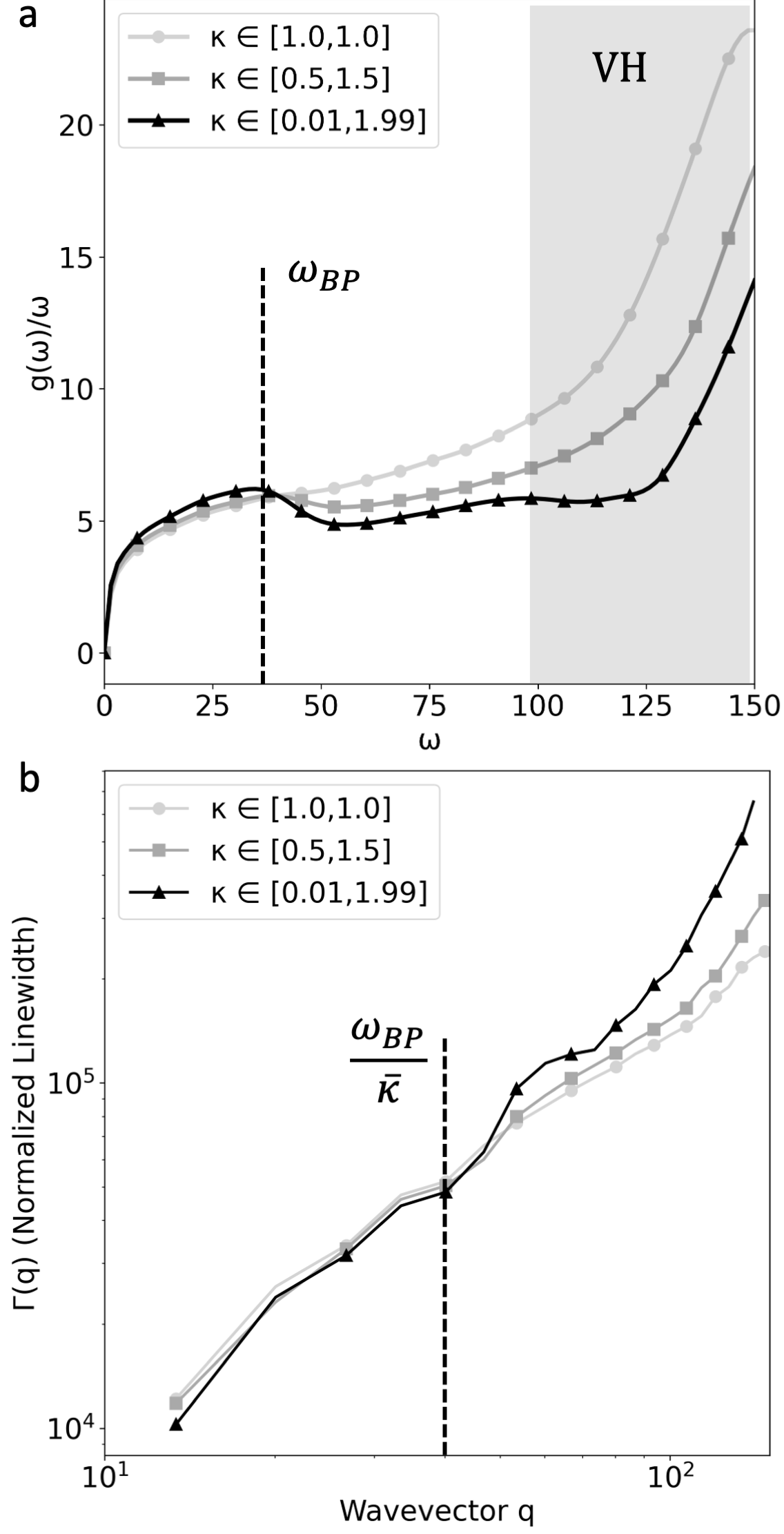}
    \caption{\textbf{a}: The reduced DOS, \(g(\omega)/\omega\), for different inhomogeneity intensities \(0\), \(0.5\), and \(0.99\) (gray, darker gray, and black, respectively), computed using Eq.~\eqref{dosq}. The \(\kappa\) distributions correspond to those in Fig.~\ref{figintensity} a--c. The vertical dashed line indicates the Boson peak (BP) position around \(\omega_{BP} \sim 40\). The gray-shaded high-frequency region (\(\omega > 100\)) corresponds to the Van Hove peak, where the dispersion relation deviates from linearity as seen in Fig.~\ref{figintensity} d--f. \textbf{b}: The reduced half-peak width \(\Gamma (q)\) of the spectral function, reflecting the flatness of \(A(\omega, q)\) at different wave vectors \(q\), as shown in Fig.~\ref{figintensity} g--i.}
    \label{dosintensity}
\end{figure}

Since the DOS is the integral of the spectral function over wave vector space (Eq.~\eqref{dosq}), the crossover at \(q \sim 1/(2L)\) induced by inhomogeneity leads to additional DOS in the low-frequency region \(\omega < \bar{\kappa}^{1/2}/(2L)\) and a reduction in DOS at high frequencies \(\omega > \bar{\kappa}^{1/2}/(2L)\), relative to the homogeneous case. The reduced DOS \(g(\omega)/\omega\) for different inhomogeneity levels is shown in Fig.~\ref{dosintensity} a. For the homogeneous case, \(g(\omega)/\omega\) exhibits a relatively flat region below \(\omega = 100\), except at very low frequencies (\(\omega < 10\)), reflecting the plane-wave nature predicted by Debye's theory. The deviations at very low frequencies are attributed to numerical precision and the causality imaginary term \(i 0^+\) in Eq.~\eqref{greenreal}, while those at high frequencies (\(\omega > 100\)) stem from the Van Hove singularity due to numerical discretization. As shown in Fig.~\ref{dosintensity} a, the introduction of inhomogeneity reduces the DOS at high frequencies (\(\omega > \bar{\kappa}^{1/2}/(2L)\)) and increases it at low frequencies (\(\omega < \bar{\kappa}^{1/2}/(2L)\)). This mechanism underlies the emergence of an anomalous peak in the reduced DOS, i.e., the Boson peak (BP).

To quantitatively assess the impact of inhomogeneity on the spectral function \(A(\omega, q)\), we define the reduced half-peak width \(\Gamma (q)\) as the full width at half maximum of \(A(\omega, q)\) at a given \(q\), divided by its height. \(\Gamma (q)\) reflects the flatness of \(A(\omega, q)\); a broader and shorter signal corresponds to a larger \(\Gamma (q)\). As shown in Fig.~\ref{dosintensity} b, \(\Gamma (q)\) is relatively smaller for inhomogeneous cases compared to the homogeneous case at small wave vectors \(q < \omega_{BP} / \bar{\kappa} \sim 1/(2L)\). Beyond this wave vector, \(\Gamma (q)\) increases rapidly for inhomogeneous cases, exceeding the homogeneous value. As an indicator of spectral flatness, \(\Gamma (q)\) reflects the scattering intensity of plane waves by the medium. Thus, elastic inhomogeneity can be interpreted as causing stronger scattering for short-wavelength excitations (large wave vectors), while long-wavelength excitations remain largely unaffected or are even enhanced by the inhomogeneity.

Two features of these results warrant special attention. First, previous particle-level simulations and experiments\cite{PhysRevLett.133.188302,Hu2022,mahajan2025flatmodeperspectivebosonpeak,Tomterud2023} attributed the origin of the low-frequency anomalous DOS (BP) to the presence of a flat dispersion relation \(\omega = \omega_{BP}\) in addition to the linear acoustic phonon dispersion \(\omega = \bar{\kappa}^{1/2} q\). However, as shown in Fig.~\ref{figintensity} e and f, such a flat dispersion relation is not observed here. Even without a flat dispersion, inhomogeneity weakens and flattens the spectral function at large wave vectors, thereby creating an anomalous DOS that violates Debye's law at low frequencies. A BP without a flat dispersion has been observed in strain glasses, where the structure is ordered but the strain is inhomogeneous in real space.\cite{Ren2021} These results suggest that the BP may have different origins in different systems. One origin involves a flat dispersion relation at the BP frequency, induced by additional vibrational modes such as 'string-like excitations'.\cite{Hu2022,PhysRevLett.133.188302,10.1063/5.0210057,ch39-6bhs,Jiang_2024} Another origin is the selective scattering of phonons due to inhomogeneity, as in inhomogeneous elastic systems.\cite{PhysRevLett.100.137402,PhysRevLett.98.025501}

\begin{figure}[htbp]
    \centering
    \includegraphics[width=\linewidth]{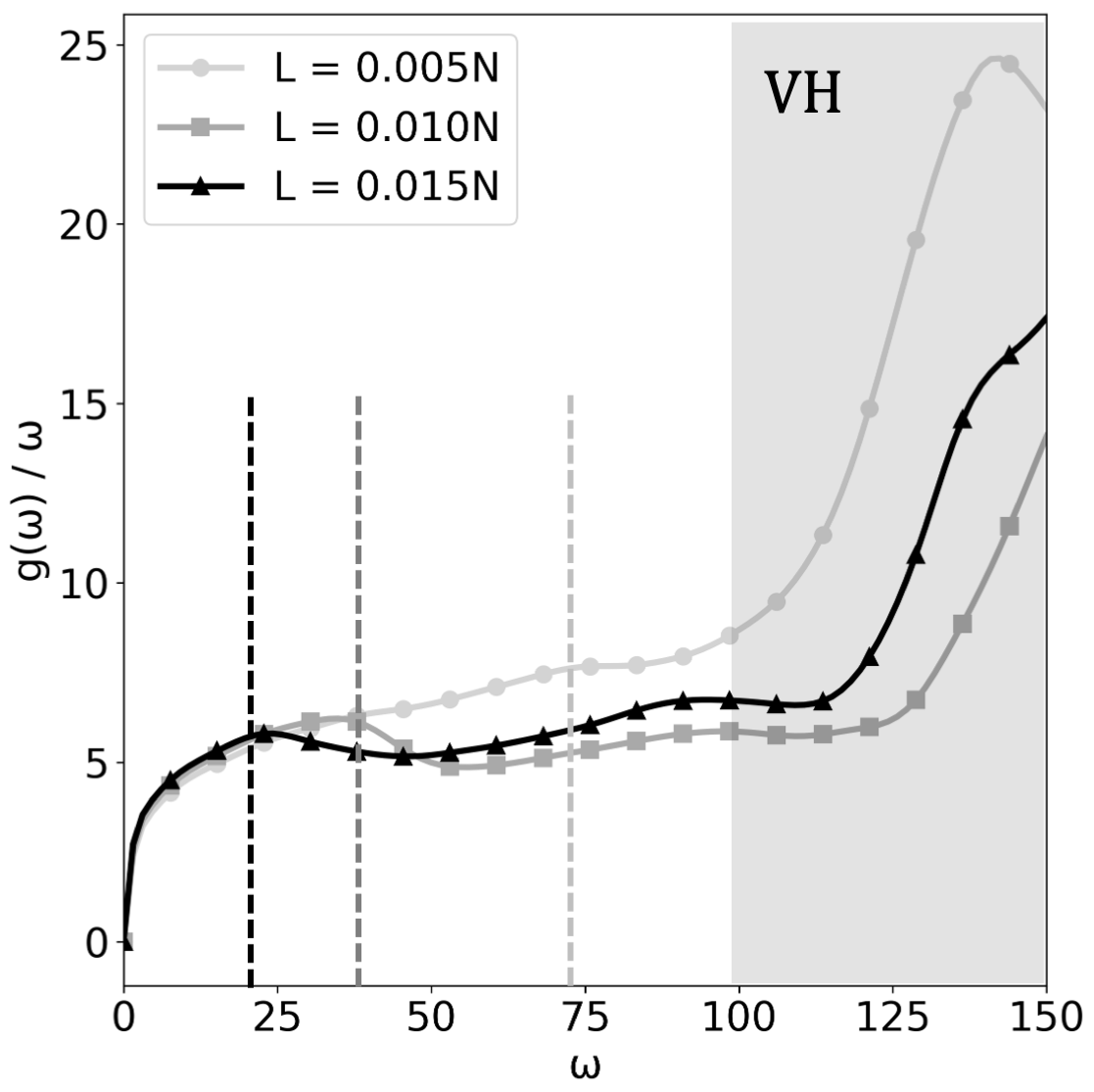}
    \caption{The reduced DOS, \(g(\omega)/\omega\), for different correlation lengths \(0.005N\), \(0.01N\), and \(0.015N\) (gray, darker gray, and black, respectively), \textbf{a}. The DOS is computed using Eq.~\eqref{dosq}, with an inhomogeneity intensity of \(0.99\). The vertical dashed lines indicate the BP positions at \(\omega_{BP} \sim 72, 40, 20\) for the respective correlation lengths.}
    \label{doslength}
\end{figure}

A second notable feature is the inverse relationship between the crossover wave vector distinguishing enhanced/suppressed spectral regions and the correlation length \(L\) of the inhomogeneity. The relation \(\omega_{BP} / \bar{\kappa}^{1/2} \sim 1/(2 L)\) has been reported in studies of string-like excitations,\cite{Hu2022,PhysRevLett.133.188302,Jiang_2024} where the BP frequency can be estimated by replacing the correlation length with the size of the string-like excitations. The DOS for different correlation lengths, shown in Fig.~\ref{doslength}, confirms this inverse relationship. As the correlation length increases, the BP position (where the reduced DOS begins to decrease) shifts to lower frequencies. For correlation lengths \(L = 0.005N, 0.01N, 0.015N\), the corresponding BP frequencies are \(\omega_{BP} = 72, 40, 20\), consistent with the inverse relation \(\omega_{BP} / \bar{\kappa}^{1/2} \sim 1/(2 L)\).

To this stage, calculations of the spectral function and DOS based on an elastically inhomogeneous medium depict a mechanism for generating low-frequency anomalous DOS that violates Debye's law. The correlation length of the inhomogeneity selects a wavelength \(2L\), and all sound waves with shorter wavelengths are strongly scattered, manifesting as a suppressed spectral function at large wave vectors \(q > 1/(2L)\). This suppression at high frequencies (\(\omega > \bar{\kappa}^{1/2}/(2L)\) consequently leads to a violation of Debye's law at low frequencies (\(\omega < \bar{\kappa}^{1/2}/(2L)\)). The correlation length dependence of the anomalous DOS aligns with recent findings on how the size of string-like excitations determines the BP frequency.\cite{Jiang_2024,10.1063/5.0210057,PhysRevLett.133.188302} However, the flat dispersion relation observed in various systems\cite{Hu2022,PhysRevLett.133.188302,Tomterud2023,mahajan2025flatmodeperspectivebosonpeak} is not captured within the inhomogeneous elasticity framework described by Eq.~\eqref{frequencyequation}. This indicates that while the inhomogeneous elasticity framework can account for the BP phenomenon,\cite{p95w-9wjc,PhysRevLett.98.025501,PhysRevLett.100.137402} its mechanism differs from the emerging picture based on flat dispersion relations. Furthermore, recent simulations and experiments have revealed string-like vibrational inhomogeneity near the BP frequency as the origin of anomalous low-frequency DOS. Eqs.~\eqref{realspectra} and \eqref{dosreal} provide the formalism to examine and visualize this vibrational inhomogeneity at different frequencies.

\begin{figure*}[htbp]
    \centering
    \includegraphics[width=\linewidth]{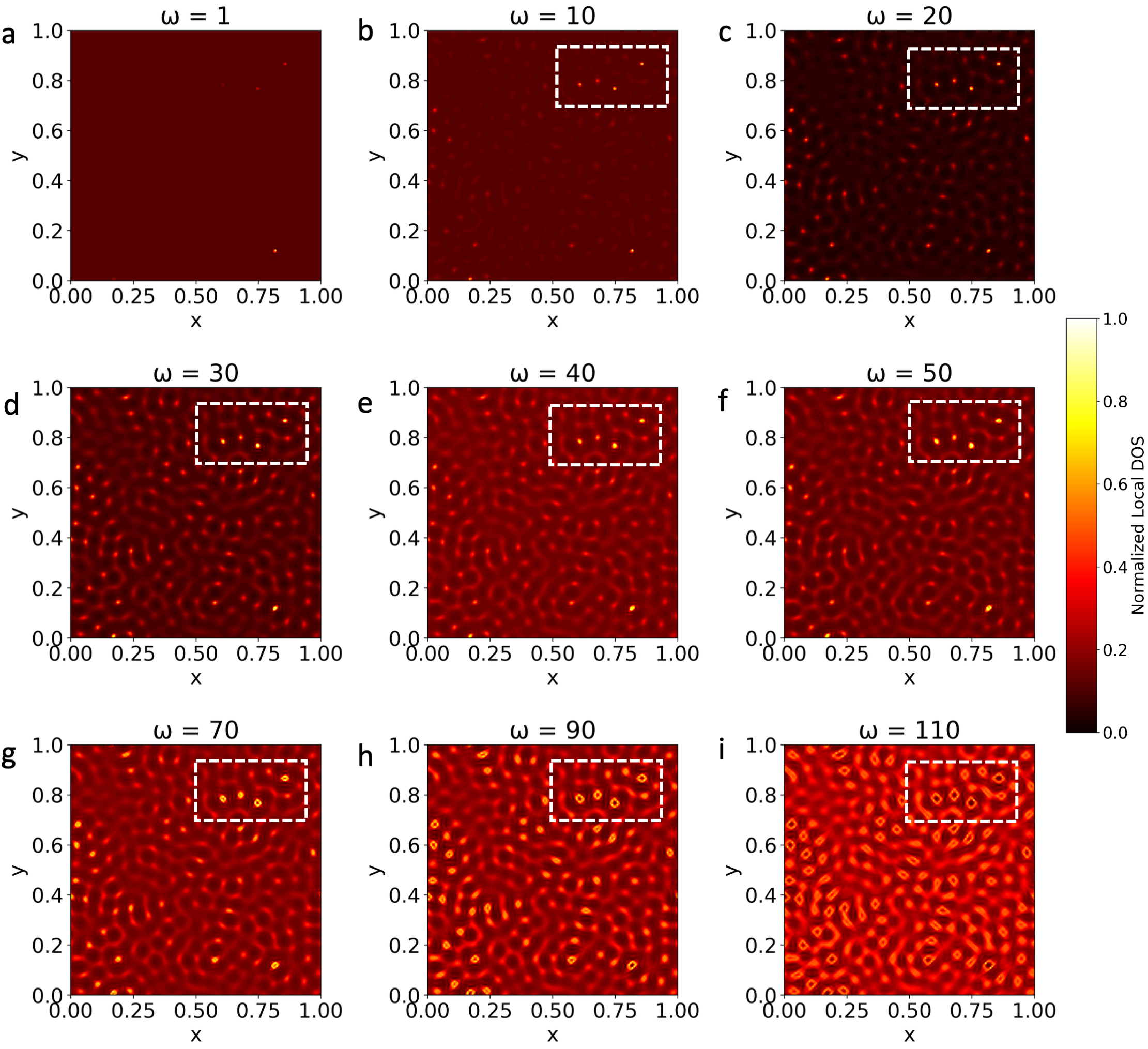}
    \caption{The local DOS \(A_{\omega} (\boldsymbol{r})\) at different frequencies \(\omega = 1, 10, 20, 30, 40, 50, 70, 90, 110\) (panels \textbf{a} to \textbf{i}). The local DOS is computed from Eqs.~\eqref{greenreal}, \eqref{realspectra}, and \eqref{dosreal}. The corresponding elastic modulus distribution \(\kappa(\boldsymbol{r})\) is shown in Fig.~\ref{figkappa}. The dashed white boxes highlight several (four) highly excited spots (regions of large local DOS) at low frequencies \(10 < \omega < 50\). Their positions correspond to soft spots (relatively small \(\kappa\)) indicated by dashed white boxes and white crosses in Fig.~\ref{figkappa}.}
    \label{figldos}
\end{figure*}

\section{Emergence of vibrational inhomogeneity}

For a homogeneous elastic system, the distribution of the local density of states (DOS), \(A_{\omega} (\boldsymbol{r}, \boldsymbol{r})\) as defined by Eqs.~\eqref{realspectra} and \eqref{dosreal}, is uniform due to translational symmetry; that is, \(A_{\omega} (\boldsymbol{r}, \boldsymbol{r}) = A_{\omega}\) depends only on frequency. Previous studies\cite{Hu2022,PhysRevLett.133.188302} have found that the local DOS (also referred to as the particle-level DOS) becomes inhomogeneous in real space, particularly near the Boson peak (BP) frequency. The term "vibrational inhomogeneity" in that context, often manifesting as string-like excitations, arises from their morphological characteristics. These findings suggest that disorder can induce vibrational inhomogeneity, although the underlying mechanism remains unclear. In the following, we demonstrate that vibrational inhomogeneity, i.e., a spatially varying local DOS, is inherently predicted by Eq.~\eqref{frequencyequation} when the elasticity is inhomogeneous.

\begin{figure}[htbp]
    \centering
    \includegraphics[width=\linewidth]{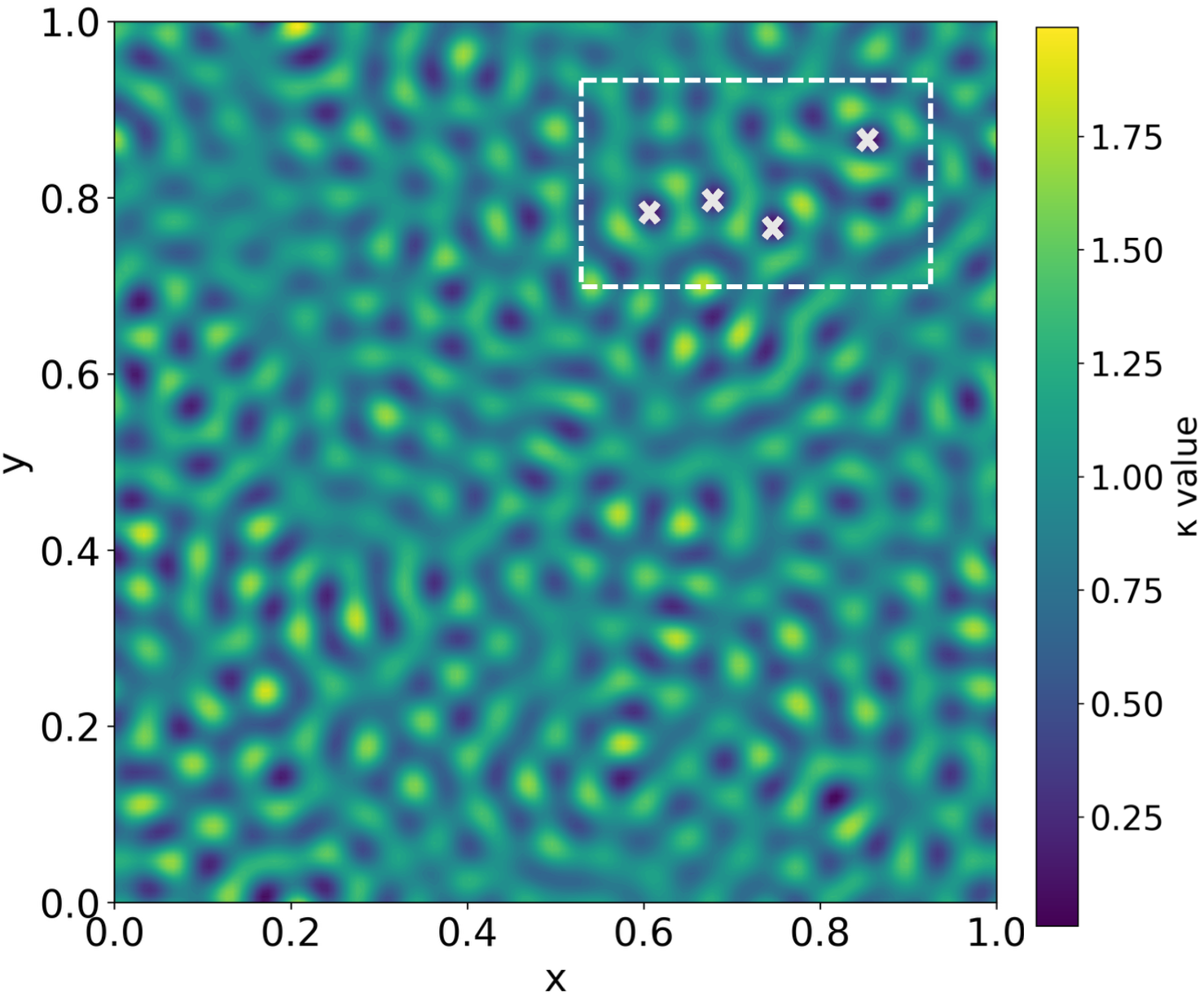}
    \caption{Distribution of the elastic modulus \(\kappa (\boldsymbol{r})\) used for calculating the local DOS shown in Fig.~\ref{figldos}. The inhomogeneity intensity is \(0.99\), and the correlation length is \(L = 0.01 N\). White crosses within the dashed white box mark the positions of highly excited spots (large local DOS) at low frequencies \(10 < \omega < 50\), as highlighted by the dashed white boxes in Fig.~\ref{figldos}.}
    \label{figkappa}
\end{figure}

The local DOS computed via Eqs.~\eqref{realspectra} and \eqref{dosreal} from low to high frequencies is shown in Fig.~\ref{figldos} a--i. The corresponding inhomogeneous elastic modulus \(\kappa (\boldsymbol{r})\) is displayed in Fig.~\ref{figkappa}, the spectral function in Fig.~\ref{dosintensity} f and i, and the total DOS in Fig.~\ref{dosintensity} (black curve). As evident in Fig.~\ref{figldos}, the local DOS is highly non-uniform at low frequencies (\(\omega < 50\)), exhibiting a few highly excited spots that contribute predominantly to the DOS in this frequency range, while other regions remain largely unexcited. Several of these highly excited spots are highlighted by white dashed boxes in Fig.~\ref{figldos}. At higher frequencies (Fig.~\ref{figldos} g--i), the highly excited spots are observed to "explode" into ring-like structures, and their dominance diminishes as other regions gradually become excited with increasing frequency. The frequency range over which these highly excited spots persist coincides with the appearance of excess low-frequency DOS, as shown by the black curve in Fig.~\ref{dosintensity}. Therefore, the highly excited spots can be identified as the origin of the excess low-frequency DOS. Their behavior is analogous to string-like excitations\cite{Hu2022,PhysRevLett.133.188302}, which are observed predominantly around the BP frequency, with the particle-level DOS becoming more uniform at higher frequencies.

To elucidate the correlation between the highly excited spots and elastic inhomogeneity, their positions (a subset thereof) are marked by white crosses within white dashed boxes in Fig.~\ref{figkappa}. It can be seen that the low frequency highly excited spots occur at very soft regions (where the elastic modulus \(\kappa\) is small) surrounded by stiff cages (where \(\kappa\) is large). As argued previously, the frequency of these highly excited spots does not depend on the local value of \(\kappa\), but is determined by the cage size and the average elastic modulus specifically, by the correlation length \(L\) and \(\bar{\kappa}\) through the relation \(\omega_{BP} \sim \bar{\kappa}^{1/2} / (2 L)\).

The emergence of highly excited spots disrupts the vibrational homogeneity characteristic of a homogeneous elastic medium at low frequencies, demonstrating that vibrational inhomogeneity is a direct consequence of elastic inhomogeneity. The computation of the local DOS in real space provides a visualization of the excess low-frequency DOS and offers a pathway to compare the theory of the inhomogeneous elastic wave equation (Eq.~\eqref{equationall}) with recent advances concerning string-like excitations. Although the specific "string" morphology is not reproduced here, the observed highly excited spots offer a clear picture for interpreting the origin and properties of vibrational inhomogeneity in particle systems, including their emergence at low frequencies and their disintegration at higher frequencies.

\section{Conclusions}
Amorphous systems characterized by inhomogeneous elasticity break the translational symmetry, which destroy the simplicity of implement theory of vibration in wave vector space. From the other side, the Green's function and spectra function constructed in real space provide direct way to study the dynamical properties of a given inhomogeneous elastic distribution. Through real space numerical analysis, it is shown that presence of elastic inhomogeneity is able to enhance the spectra function under low frequency but suppress it under high frequency, and therefore, the anomalous total DOS violating Debye's law appear around the frequency that distinguish low and high. In addition, it is confirmed that the special frequency is determined by correlation length of elastic inhomogeneity, the property which agree with recent particle level foundings. Furthermore, the emergence of vibrational inhomogeneity, characterizing as highly excited spots, is demonstrated to be result directly of inhomogeneous elasticity. The numerical results in this work provide a picture about the origin of anomalous DOS observed in amorphous solids, and bridge the continue theory with recent progresses of particle level vibrational inhomogeneity.

%However, the recent proposal of flat dispersion relation to be the origin of BP,\cite{Hu2022,mahajan2025flatmodeperspectivebosonpeak} is not able to be included in the theoretical framework of inhomogeneous elasticity, neither wave vector space approach,\cite{PhysRevLett.98.025501,PhysRevLett.100.137402} nor the real space approach in this work. This indicate that statical elastic inhomogeneity as in Eq.\eqref{frequencyequation} may not able to grasp the full picture of amorphous solids. 

\section{Acknowledgments}
%The author would like to thank Matteo Baggioli and Bingyu Cui for discussions and suggestions on this work.

%\bibliographystyle{apsrev4-1}
%\bibliography{main}

\begin{thebibliography}{36}
\expandafter\ifx\csname natexlab\endcsname\relax\def\natexlab#1{#1}\fi
\expandafter\ifx\csname bibnamefont\endcsname\relax
  \def\bibnamefont#1{#1}\fi
\expandafter\ifx\csname bibfnamefont\endcsname\relax
  \def\bibfnamefont#1{#1}\fi
\expandafter\ifx\csname citenamefont\endcsname\relax
  \def\citenamefont#1{#1}\fi
\expandafter\ifx\csname url\endcsname\relax
  \def\url#1{\texttt{#1}}\fi
\expandafter\ifx\csname urlprefix\endcsname\relax\def\urlprefix{URL }\fi
\providecommand{\bibinfo}[2]{#2}
\providecommand{\eprint}[2][]{\url{#2}}

\bibitem[{\citenamefont{Debye}(1912)}]{debye}
\bibinfo{author}{\bibfnamefont{P.}~\bibnamefont{Debye}}, \bibinfo{journal}{Annalen der Physik} \textbf{\bibinfo{volume}{344}}, \bibinfo{pages}{789} (\bibinfo{year}{1912}).

\bibitem[{\citenamefont{Zeller and Pohl}(1971)}]{PhysRevB.4.2029}
\bibinfo{author}{\bibfnamefont{R.~C.} \bibnamefont{Zeller}} \bibnamefont{and} \bibinfo{author}{\bibfnamefont{R.~O.} \bibnamefont{Pohl}}, \bibinfo{journal}{Phys. Rev. B} \textbf{\bibinfo{volume}{4}}, \bibinfo{pages}{2029} (\bibinfo{year}{1971}), \urlprefix\url{https://link.aps.org/doi/10.1103/PhysRevB.4.2029}.

\bibitem[{\citenamefont{P\'erez-Casta\~neda et~al.}(2014)\citenamefont{P\'erez-Casta\~neda, Jim\'enez-Riob\'oo, and Ramos}}]{PhysRevLett.112.165901}
\bibinfo{author}{\bibfnamefont{T.}~\bibnamefont{P\'erez-Casta\~neda}}, \bibinfo{author}{\bibfnamefont{R.~J.} \bibnamefont{Jim\'enez-Riob\'oo}}, \bibnamefont{and} \bibinfo{author}{\bibfnamefont{M.~A.} \bibnamefont{Ramos}}, \bibinfo{journal}{Phys. Rev. Lett.} \textbf{\bibinfo{volume}{112}}, \bibinfo{pages}{165901} (\bibinfo{year}{2014}), \urlprefix\url{https://link.aps.org/doi/10.1103/PhysRevLett.112.165901}.

\bibitem[{\citenamefont{Gurevich et~al.}(2003)\citenamefont{Gurevich, Parshin, and Schober}}]{PhysRevB.67.094203}
\bibinfo{author}{\bibfnamefont{V.~L.} \bibnamefont{Gurevich}}, \bibinfo{author}{\bibfnamefont{D.~A.} \bibnamefont{Parshin}}, \bibnamefont{and} \bibinfo{author}{\bibfnamefont{H.~R.} \bibnamefont{Schober}}, \bibinfo{journal}{Phys. Rev. B} \textbf{\bibinfo{volume}{67}}, \bibinfo{pages}{094203} (\bibinfo{year}{2003}), \urlprefix\url{https://link.aps.org/doi/10.1103/PhysRevB.67.094203}.

\bibitem[{\citenamefont{Jiang et~al.}(2024{\natexlab{a}})\citenamefont{Jiang, Zheng, Chen, Baggioli, and Zhang}}]{PhysRevLett.133.188302}
\bibinfo{author}{\bibfnamefont{C.}~\bibnamefont{Jiang}}, \bibinfo{author}{\bibfnamefont{Z.}~\bibnamefont{Zheng}}, \bibinfo{author}{\bibfnamefont{Y.}~\bibnamefont{Chen}}, \bibinfo{author}{\bibfnamefont{M.}~\bibnamefont{Baggioli}}, \bibnamefont{and} \bibinfo{author}{\bibfnamefont{J.}~\bibnamefont{Zhang}}, \bibinfo{journal}{Phys. Rev. Lett.} \textbf{\bibinfo{volume}{133}}, \bibinfo{pages}{188302} (\bibinfo{year}{2024}{\natexlab{a}}), \urlprefix\url{https://link.aps.org/doi/10.1103/PhysRevLett.133.188302}.

\bibitem[{\citenamefont{Moriel et~al.}(2024)\citenamefont{Moriel, Lerner, and Bouchbinder}}]{PhysRevResearch.6.023053}
\bibinfo{author}{\bibfnamefont{A.}~\bibnamefont{Moriel}}, \bibinfo{author}{\bibfnamefont{E.}~\bibnamefont{Lerner}}, \bibnamefont{and} \bibinfo{author}{\bibfnamefont{E.}~\bibnamefont{Bouchbinder}}, \bibinfo{journal}{Phys. Rev. Res.} \textbf{\bibinfo{volume}{6}}, \bibinfo{pages}{023053} (\bibinfo{year}{2024}), \urlprefix\url{https://link.aps.org/doi/10.1103/PhysRevResearch.6.023053}.

\bibitem[{\citenamefont{Lerner and Bouchbinder}(2023)}]{Edan2023}
\bibinfo{author}{\bibfnamefont{E.}~\bibnamefont{Lerner}} \bibnamefont{and} \bibinfo{author}{\bibfnamefont{E.}~\bibnamefont{Bouchbinder}}, \bibinfo{journal}{The Journal of Chemical Physics} \textbf{\bibinfo{volume}{158}}, \bibinfo{pages}{194503} (\bibinfo{year}{2023}), ISSN \bibinfo{issn}{0021-9606}, \urlprefix\url{https://doi.org/10.1063/5.0147889}.

\bibitem[{\citenamefont{Gonz{\'a}lez-Jim{\'e}nez et~al.}(2023)\citenamefont{Gonz{\'a}lez-Jim{\'e}nez, Barnard, Russell, Tukachev, Javornik, Hayes, Farrell, Guinane, Senn, Smith et~al.}}]{Gj2023}
\bibinfo{author}{\bibfnamefont{M.}~\bibnamefont{Gonz{\'a}lez-Jim{\'e}nez}}, \bibinfo{author}{\bibfnamefont{T.}~\bibnamefont{Barnard}}, \bibinfo{author}{\bibfnamefont{B.~A.} \bibnamefont{Russell}}, \bibinfo{author}{\bibfnamefont{N.~V.} \bibnamefont{Tukachev}}, \bibinfo{author}{\bibfnamefont{U.}~\bibnamefont{Javornik}}, \bibinfo{author}{\bibfnamefont{L.-A.} \bibnamefont{Hayes}}, \bibinfo{author}{\bibfnamefont{A.~J.} \bibnamefont{Farrell}}, \bibinfo{author}{\bibfnamefont{S.}~\bibnamefont{Guinane}}, \bibinfo{author}{\bibfnamefont{H.~M.} \bibnamefont{Senn}}, \bibinfo{author}{\bibfnamefont{A.~J.} \bibnamefont{Smith}}, \bibnamefont{et~al.}, \bibinfo{journal}{Nature Communications} \textbf{\bibinfo{volume}{14}}, \bibinfo{pages}{215} (\bibinfo{year}{2023}), ISSN \bibinfo{issn}{2041-1723}, \urlprefix\url{https://doi.org/10.1038/s41467-023-35878-6}.

\bibitem[{\citenamefont{Schmid and Schirmacher}(2008)}]{PhysRevLett.100.137402}
\bibinfo{author}{\bibfnamefont{B.}~\bibnamefont{Schmid}} \bibnamefont{and} \bibinfo{author}{\bibfnamefont{W.}~\bibnamefont{Schirmacher}}, \bibinfo{journal}{Phys. Rev. Lett.} \textbf{\bibinfo{volume}{100}}, \bibinfo{pages}{137402} (\bibinfo{year}{2008}), \urlprefix\url{https://link.aps.org/doi/10.1103/PhysRevLett.100.137402}.

\bibitem[{\citenamefont{Schirmacher et~al.}(2007)\citenamefont{Schirmacher, Ruocco, and Scopigno}}]{PhysRevLett.98.025501}
\bibinfo{author}{\bibfnamefont{W.}~\bibnamefont{Schirmacher}}, \bibinfo{author}{\bibfnamefont{G.}~\bibnamefont{Ruocco}}, \bibnamefont{and} \bibinfo{author}{\bibfnamefont{T.}~\bibnamefont{Scopigno}}, \bibinfo{journal}{Phys. Rev. Lett.} \textbf{\bibinfo{volume}{98}}, \bibinfo{pages}{025501} (\bibinfo{year}{2007}), \urlprefix\url{https://link.aps.org/doi/10.1103/PhysRevLett.98.025501}.

\bibitem[{\citenamefont{Caroli and Lema\^{\i}tre}(2019)}]{PhysRevLett.123.055501}
\bibinfo{author}{\bibfnamefont{C.}~\bibnamefont{Caroli}} \bibnamefont{and} \bibinfo{author}{\bibfnamefont{A.}~\bibnamefont{Lema\^{\i}tre}}, \bibinfo{journal}{Phys. Rev. Lett.} \textbf{\bibinfo{volume}{123}}, \bibinfo{pages}{055501} (\bibinfo{year}{2019}), \urlprefix\url{https://link.aps.org/doi/10.1103/PhysRevLett.123.055501}.

\bibitem[{\citenamefont{Vogel and Fuchs}(2023)}]{PhysRevLett.130.236101}
\bibinfo{author}{\bibfnamefont{F.}~\bibnamefont{Vogel}} \bibnamefont{and} \bibinfo{author}{\bibfnamefont{M.}~\bibnamefont{Fuchs}}, \bibinfo{journal}{Phys. Rev. Lett.} \textbf{\bibinfo{volume}{130}}, \bibinfo{pages}{236101} (\bibinfo{year}{2023}), \urlprefix\url{https://link.aps.org/doi/10.1103/PhysRevLett.130.236101}.

\bibitem[{\citenamefont{Shintani and Tanaka}(2008)}]{Shintani2008}
\bibinfo{author}{\bibfnamefont{H.}~\bibnamefont{Shintani}} \bibnamefont{and} \bibinfo{author}{\bibfnamefont{H.}~\bibnamefont{Tanaka}}, \bibinfo{journal}{Nature Materials} \textbf{\bibinfo{volume}{7}}, \bibinfo{pages}{870} (\bibinfo{year}{2008}), ISSN \bibinfo{issn}{1476-4660}, \urlprefix\url{https://doi.org/10.1038/nmat2293}.

\bibitem[{\citenamefont{Tong et~al.}(2015)\citenamefont{Tong, Tan, and Xu}}]{Tong2015}
\bibinfo{author}{\bibfnamefont{H.}~\bibnamefont{Tong}}, \bibinfo{author}{\bibfnamefont{P.}~\bibnamefont{Tan}}, \bibnamefont{and} \bibinfo{author}{\bibfnamefont{N.}~\bibnamefont{Xu}}, \bibinfo{journal}{Scientific Reports} \textbf{\bibinfo{volume}{5}}, \bibinfo{pages}{15378} (\bibinfo{year}{2015}), ISSN \bibinfo{issn}{2045-2322}, \urlprefix\url{https://doi.org/10.1038/srep15378}.

\bibitem[{\citenamefont{Hu and Tanaka}(2022)}]{Hu2022}
\bibinfo{author}{\bibfnamefont{Y.-C.} \bibnamefont{Hu}} \bibnamefont{and} \bibinfo{author}{\bibfnamefont{H.}~\bibnamefont{Tanaka}}, \bibinfo{journal}{Nature Physics} \textbf{\bibinfo{volume}{18}}, \bibinfo{pages}{669} (\bibinfo{year}{2022}), ISSN \bibinfo{issn}{1745-2481}, \urlprefix\url{https://doi.org/10.1038/s41567-022-01628-6}.

\bibitem[{\citenamefont{Mahajan and Ciamarra}(2021)}]{PhysRevLett.127.215504}
\bibinfo{author}{\bibfnamefont{S.}~\bibnamefont{Mahajan}} \bibnamefont{and} \bibinfo{author}{\bibfnamefont{M.~P.} \bibnamefont{Ciamarra}}, \bibinfo{journal}{Phys. Rev. Lett.} \textbf{\bibinfo{volume}{127}}, \bibinfo{pages}{215504} (\bibinfo{year}{2021}), \urlprefix\url{https://link.aps.org/doi/10.1103/PhysRevLett.127.215504}.

\bibitem[{\citenamefont{Zhang et~al.}(2013)\citenamefont{Zhang, Khalkhali, Liu, and Douglas}}]{10.1063/1.4769267}
\bibinfo{author}{\bibfnamefont{H.}~\bibnamefont{Zhang}}, \bibinfo{author}{\bibfnamefont{M.}~\bibnamefont{Khalkhali}}, \bibinfo{author}{\bibfnamefont{Q.}~\bibnamefont{Liu}}, \bibnamefont{and} \bibinfo{author}{\bibfnamefont{J.~F.} \bibnamefont{Douglas}}, \bibinfo{journal}{The Journal of Chemical Physics} \textbf{\bibinfo{volume}{138}}, \bibinfo{pages}{12A538} (\bibinfo{year}{2013}), ISSN \bibinfo{issn}{0021-9606}, \urlprefix\url{https://doi.org/10.1063/1.4769267}.

\bibitem[{\citenamefont{Zhang et~al.}(2011)\citenamefont{Zhang, Kalvapalle, and Douglas}}]{Zhang2011}
\bibinfo{author}{\bibfnamefont{H.}~\bibnamefont{Zhang}}, \bibinfo{author}{\bibfnamefont{P.}~\bibnamefont{Kalvapalle}}, \bibnamefont{and} \bibinfo{author}{\bibfnamefont{J.~F.} \bibnamefont{Douglas}}, \bibinfo{journal}{The Journal of Physical Chemistry B} \textbf{\bibinfo{volume}{115}}, \bibinfo{pages}{14068} (\bibinfo{year}{2011}), ISSN \bibinfo{issn}{1520-6106}, \urlprefix\url{https://doi.org/10.1021/jp203765x}.

\bibitem[{\citenamefont{Wang et~al.}(2019)\citenamefont{Wang, Ninarello, Guan, Berthier, Szamel, and Flenner}}]{Wang2019}
\bibinfo{author}{\bibfnamefont{L.}~\bibnamefont{Wang}}, \bibinfo{author}{\bibfnamefont{A.}~\bibnamefont{Ninarello}}, \bibinfo{author}{\bibfnamefont{P.}~\bibnamefont{Guan}}, \bibinfo{author}{\bibfnamefont{L.}~\bibnamefont{Berthier}}, \bibinfo{author}{\bibfnamefont{G.}~\bibnamefont{Szamel}}, \bibnamefont{and} \bibinfo{author}{\bibfnamefont{E.}~\bibnamefont{Flenner}}, \bibinfo{journal}{Nature Communications} \textbf{\bibinfo{volume}{10}}, \bibinfo{pages}{26} (\bibinfo{year}{2019}), ISSN \bibinfo{issn}{2041-1723}, \urlprefix\url{https://doi.org/10.1038/s41467-018-07978-1}.

\bibitem[{\citenamefont{Rainone et~al.}(2020)\citenamefont{Rainone, Bouchbinder, and Lerner}}]{Corrado2020}
\bibinfo{author}{\bibfnamefont{C.}~\bibnamefont{Rainone}}, \bibinfo{author}{\bibfnamefont{E.}~\bibnamefont{Bouchbinder}}, \bibnamefont{and} \bibinfo{author}{\bibfnamefont{E.}~\bibnamefont{Lerner}}, \bibinfo{journal}{Proceedings of the National Academy of Sciences} \textbf{\bibinfo{volume}{117}}, \bibinfo{pages}{5228} (\bibinfo{year}{2020}), \eprint{https://www.pnas.org/doi/pdf/10.1073/pnas.1919958117}, \urlprefix\url{https://www.pnas.org/doi/abs/10.1073/pnas.1919958117}.

\bibitem[{\citenamefont{Mahajan and Ciamarra}(2023)}]{10.21468/SciPostPhys.15.2.069}
\bibinfo{author}{\bibfnamefont{S.}~\bibnamefont{Mahajan}} \bibnamefont{and} \bibinfo{author}{\bibfnamefont{M.~P.} \bibnamefont{Ciamarra}}, \bibinfo{journal}{SciPost Phys.} \textbf{\bibinfo{volume}{15}}, \bibinfo{pages}{069} (\bibinfo{year}{2023}), \urlprefix\url{https://scipost.org/10.21468/SciPostPhys.15.2.069}.

\bibitem[{\citenamefont{Xu et~al.}(2024)\citenamefont{Xu, Zhang, Tong, Wang, and Xu}}]{Xu2024}
\bibinfo{author}{\bibfnamefont{D.}~\bibnamefont{Xu}}, \bibinfo{author}{\bibfnamefont{S.}~\bibnamefont{Zhang}}, \bibinfo{author}{\bibfnamefont{H.}~\bibnamefont{Tong}}, \bibinfo{author}{\bibfnamefont{L.}~\bibnamefont{Wang}}, \bibnamefont{and} \bibinfo{author}{\bibfnamefont{N.}~\bibnamefont{Xu}}, \bibinfo{journal}{Nature Communications} \textbf{\bibinfo{volume}{15}}, \bibinfo{pages}{1424} (\bibinfo{year}{2024}), ISSN \bibinfo{issn}{2041-1723}, \urlprefix\url{https://doi.org/10.1038/s41467-024-45671-8}.

\bibitem[{\citenamefont{Wang et~al.}(2018)\citenamefont{Wang, Hong, Wang, Schirmacher, and Zhang}}]{PhysRevB.98.174207}
\bibinfo{author}{\bibfnamefont{Y.}~\bibnamefont{Wang}}, \bibinfo{author}{\bibfnamefont{L.}~\bibnamefont{Hong}}, \bibinfo{author}{\bibfnamefont{Y.}~\bibnamefont{Wang}}, \bibinfo{author}{\bibfnamefont{W.}~\bibnamefont{Schirmacher}}, \bibnamefont{and} \bibinfo{author}{\bibfnamefont{J.}~\bibnamefont{Zhang}}, \bibinfo{journal}{Phys. Rev. B} \textbf{\bibinfo{volume}{98}}, \bibinfo{pages}{174207} (\bibinfo{year}{2018}), \urlprefix\url{https://link.aps.org/doi/10.1103/PhysRevB.98.174207}.

\bibitem[{\citenamefont{Zhang et~al.}(2017)\citenamefont{Zhang, Zheng, Wang, Zhang, Jin, Hong, Wang, and Zhang}}]{Zhang2017}
\bibinfo{author}{\bibfnamefont{L.}~\bibnamefont{Zhang}}, \bibinfo{author}{\bibfnamefont{J.}~\bibnamefont{Zheng}}, \bibinfo{author}{\bibfnamefont{Y.}~\bibnamefont{Wang}}, \bibinfo{author}{\bibfnamefont{L.}~\bibnamefont{Zhang}}, \bibinfo{author}{\bibfnamefont{Z.}~\bibnamefont{Jin}}, \bibinfo{author}{\bibfnamefont{L.}~\bibnamefont{Hong}}, \bibinfo{author}{\bibfnamefont{Y.}~\bibnamefont{Wang}}, \bibnamefont{and} \bibinfo{author}{\bibfnamefont{J.}~\bibnamefont{Zhang}}, \bibinfo{journal}{Nature Communications} \textbf{\bibinfo{volume}{8}}, \bibinfo{pages}{67} (\bibinfo{year}{2017}), ISSN \bibinfo{issn}{2041-1723}, \urlprefix\url{https://doi.org/10.1038/s41467-017-00106-5}.

\bibitem[{\citenamefont{Sun et~al.}(2025)\citenamefont{Sun, Chen, Ji, Zhou, Tong, Chen, Chen, Tanaka, and Tan}}]{sun2025disentanglinghierarchicalrelaxationsglass}
\bibinfo{author}{\bibfnamefont{W.}~\bibnamefont{Sun}}, \bibinfo{author}{\bibfnamefont{Y.}~\bibnamefont{Chen}}, \bibinfo{author}{\bibfnamefont{W.}~\bibnamefont{Ji}}, \bibinfo{author}{\bibfnamefont{Y.}~\bibnamefont{Zhou}}, \bibinfo{author}{\bibfnamefont{H.}~\bibnamefont{Tong}}, \bibinfo{author}{\bibfnamefont{K.}~\bibnamefont{Chen}}, \bibinfo{author}{\bibfnamefont{X.}~\bibnamefont{Chen}}, \bibinfo{author}{\bibfnamefont{H.}~\bibnamefont{Tanaka}}, \bibnamefont{and} \bibinfo{author}{\bibfnamefont{P.}~\bibnamefont{Tan}}, \emph{\bibinfo{title}{Disentangling hierarchical relaxations in glass formers via dynamic eigenmodes}} (\bibinfo{year}{2025}), \eprint{2505.19934}, \urlprefix\url{https://arxiv.org/abs/2505.19934}.

\bibitem[{\citenamefont{Tan et~al.}(2012)\citenamefont{Tan, Xu, Schofield, and Xu}}]{PhysRevLett.108.095501}
\bibinfo{author}{\bibfnamefont{P.}~\bibnamefont{Tan}}, \bibinfo{author}{\bibfnamefont{N.}~\bibnamefont{Xu}}, \bibinfo{author}{\bibfnamefont{A.~B.} \bibnamefont{Schofield}}, \bibnamefont{and} \bibinfo{author}{\bibfnamefont{L.}~\bibnamefont{Xu}}, \bibinfo{journal}{Phys. Rev. Lett.} \textbf{\bibinfo{volume}{108}}, \bibinfo{pages}{095501} (\bibinfo{year}{2012}), \urlprefix\url{https://link.aps.org/doi/10.1103/PhysRevLett.108.095501}.

\bibitem[{\citenamefont{Jiang and Baggioli}(2024)}]{Jiang_2024}
\bibinfo{author}{\bibfnamefont{C.}~\bibnamefont{Jiang}} \bibnamefont{and} \bibinfo{author}{\bibfnamefont{M.}~\bibnamefont{Baggioli}}, \bibinfo{journal}{Journal of Physics: Condensed Matter} \textbf{\bibinfo{volume}{36}}, \bibinfo{pages}{505101} (\bibinfo{year}{2024}), \urlprefix\url{https://doi.org/10.1088/1361-648X/ad789c}.

\bibitem[{\citenamefont{Bianchi et~al.}(2020)\citenamefont{Bianchi, Giordano, and Lund}}]{PhysRevB.101.174311}
\bibinfo{author}{\bibfnamefont{E.}~\bibnamefont{Bianchi}}, \bibinfo{author}{\bibfnamefont{V.~M.} \bibnamefont{Giordano}}, \bibnamefont{and} \bibinfo{author}{\bibfnamefont{F.}~\bibnamefont{Lund}}, \bibinfo{journal}{Phys. Rev. B} \textbf{\bibinfo{volume}{101}}, \bibinfo{pages}{174311} (\bibinfo{year}{2020}), \urlprefix\url{https://link.aps.org/doi/10.1103/PhysRevB.101.174311}.

\bibitem[{\citenamefont{Jiang}(2025)}]{Jiang_2025}
\bibinfo{author}{\bibfnamefont{C.}~\bibnamefont{Jiang}}, \bibinfo{journal}{Journal of Physics: Condensed Matter} \textbf{\bibinfo{volume}{37}}, \bibinfo{pages}{305401} (\bibinfo{year}{2025}), \urlprefix\url{https://doi.org/10.1088/1361-648X/adf0d3}.

\bibitem[{\citenamefont{Baggioli and Zaccone}(2019)}]{PhysRevLett.122.145501}
\bibinfo{author}{\bibfnamefont{M.}~\bibnamefont{Baggioli}} \bibnamefont{and} \bibinfo{author}{\bibfnamefont{A.}~\bibnamefont{Zaccone}}, \bibinfo{journal}{Phys. Rev. Lett.} \textbf{\bibinfo{volume}{122}}, \bibinfo{pages}{145501} (\bibinfo{year}{2019}), \urlprefix\url{https://link.aps.org/doi/10.1103/PhysRevLett.122.145501}.

\bibitem[{\citenamefont{Mahajan et~al.}(2025{\natexlab{a}})\citenamefont{Mahajan, Huang, Jiang, Wang, Ciamarra, Zhang, and Baggioli}}]{mahajan2025flatmodeperspectivebosonpeak}
\bibinfo{author}{\bibfnamefont{S.}~\bibnamefont{Mahajan}}, \bibinfo{author}{\bibfnamefont{L.-Z.} \bibnamefont{Huang}}, \bibinfo{author}{\bibfnamefont{C.}~\bibnamefont{Jiang}}, \bibinfo{author}{\bibfnamefont{Y.-J.} \bibnamefont{Wang}}, \bibinfo{author}{\bibfnamefont{M.~P.} \bibnamefont{Ciamarra}}, \bibinfo{author}{\bibfnamefont{J.}~\bibnamefont{Zhang}}, \bibnamefont{and} \bibinfo{author}{\bibfnamefont{M.}~\bibnamefont{Baggioli}}, \emph{\bibinfo{title}{A flat-mode perspective on the boson peak in amorphous solids}} (\bibinfo{year}{2025}{\natexlab{a}}), \eprint{2509.06340}, \urlprefix\url{https://arxiv.org/abs/2509.06340}.

\bibitem[{\citenamefont{Tomterud et~al.}(2023)\citenamefont{Tomterud, Eder, B{\"u}chner, Wondraczek, Simonsen, Schirmacher, Manson, and Holst}}]{Tomterud2023}
\bibinfo{author}{\bibfnamefont{M.}~\bibnamefont{Tomterud}}, \bibinfo{author}{\bibfnamefont{S.~D.} \bibnamefont{Eder}}, \bibinfo{author}{\bibfnamefont{C.}~\bibnamefont{B{\"u}chner}}, \bibinfo{author}{\bibfnamefont{L.}~\bibnamefont{Wondraczek}}, \bibinfo{author}{\bibfnamefont{I.}~\bibnamefont{Simonsen}}, \bibinfo{author}{\bibfnamefont{W.}~\bibnamefont{Schirmacher}}, \bibinfo{author}{\bibfnamefont{J.~R.} \bibnamefont{Manson}}, \bibnamefont{and} \bibinfo{author}{\bibfnamefont{B.}~\bibnamefont{Holst}}, \bibinfo{journal}{Nature Physics} \textbf{\bibinfo{volume}{19}}, \bibinfo{pages}{1910} (\bibinfo{year}{2023}), ISSN \bibinfo{issn}{1745-2481}, \urlprefix\url{https://doi.org/10.1038/s41567-023-02177-2}.

\bibitem[{\citenamefont{Ren et~al.}(2021)\citenamefont{Ren, Zong, Tao, Sun, Sun, Xue, Ding, and Wang}}]{Ren2021}
\bibinfo{author}{\bibfnamefont{S.}~\bibnamefont{Ren}}, \bibinfo{author}{\bibfnamefont{H.-X.} \bibnamefont{Zong}}, \bibinfo{author}{\bibfnamefont{X.-F.} \bibnamefont{Tao}}, \bibinfo{author}{\bibfnamefont{Y.-H.} \bibnamefont{Sun}}, \bibinfo{author}{\bibfnamefont{B.-A.} \bibnamefont{Sun}}, \bibinfo{author}{\bibfnamefont{D.-Z.} \bibnamefont{Xue}}, \bibinfo{author}{\bibfnamefont{X.-D.} \bibnamefont{Ding}}, \bibnamefont{and} \bibinfo{author}{\bibfnamefont{W.-H.} \bibnamefont{Wang}}, \bibinfo{journal}{Nature Communications} \textbf{\bibinfo{volume}{12}}, \bibinfo{pages}{5755} (\bibinfo{year}{2021}), ISSN \bibinfo{issn}{2041-1723}, \urlprefix\url{https://doi.org/10.1038/s41467-021-26029-w}.

\bibitem[{\citenamefont{Jiang et~al.}(2024{\natexlab{b}})\citenamefont{Jiang, Baggioli, and Douglas}}]{10.1063/5.0210057}
\bibinfo{author}{\bibfnamefont{C.}~\bibnamefont{Jiang}}, \bibinfo{author}{\bibfnamefont{M.}~\bibnamefont{Baggioli}}, \bibnamefont{and} \bibinfo{author}{\bibfnamefont{J.~F.} \bibnamefont{Douglas}}, \bibinfo{journal}{The Journal of Chemical Physics} \textbf{\bibinfo{volume}{160}}, \bibinfo{pages}{214505} (\bibinfo{year}{2024}{\natexlab{b}}), ISSN \bibinfo{issn}{0021-9606}, \urlprefix\url{https://doi.org/10.1063/5.0210057}.

\bibitem[{\citenamefont{Mahajan et~al.}(2025{\natexlab{b}})\citenamefont{Mahajan, Seow Yang~Han, Jiang, Baggioli, and Ciamarra}}]{ch39-6bhs}
\bibinfo{author}{\bibfnamefont{S.}~\bibnamefont{Mahajan}}, \bibinfo{author}{\bibfnamefont{D.}~\bibnamefont{Seow Yang~Han}}, \bibinfo{author}{\bibfnamefont{C.}~\bibnamefont{Jiang}}, \bibinfo{author}{\bibfnamefont{M.}~\bibnamefont{Baggioli}}, \bibnamefont{and} \bibinfo{author}{\bibfnamefont{M.~P.} \bibnamefont{Ciamarra}}, \bibinfo{journal}{Phys. Rev. E} \textbf{\bibinfo{volume}{112}}, \bibinfo{pages}{035413} (\bibinfo{year}{2025}{\natexlab{b}}), \urlprefix\url{https://link.aps.org/doi/10.1103/ch39-6bhs}.

\bibitem[{\citenamefont{Schirmacher and Ruocco}(2025)}]{p95w-9wjc}
\bibinfo{author}{\bibfnamefont{W.}~\bibnamefont{Schirmacher}} \bibnamefont{and} \bibinfo{author}{\bibfnamefont{G.}~\bibnamefont{Ruocco}}, \bibinfo{journal}{Phys. Rev. Lett.} \textbf{\bibinfo{volume}{135}}, \bibinfo{pages}{126102} (\bibinfo{year}{2025}), \urlprefix\url{https://link.aps.org/doi/10.1103/p95w-9wjc}.

\end{thebibliography}

\end{document}